\documentclass[english]{extarticle}

\usepackage[T1]{fontenc}
\usepackage[latin9]{inputenc}
\usepackage{geometry}
\usepackage{color}
\usepackage{babel}
\usepackage{textcomp}
\usepackage{mathrsfs}
\usepackage{amsmath}
\usepackage{amsthm}
\usepackage{amssymb}
\usepackage{stackrel}
\usepackage{setspace}
\usepackage[unicode=true,pdfusetitle,
 bookmarks=true,bookmarksnumbered=false,bookmarksopen=false,
 breaklinks=false,pdfborder={0 0 1},backref=false,colorlinks=blue]
 {hyperref}

\makeatletter

\providecommand{\tabularnewline}{\\}

\numberwithin{figure}{section}
\numberwithin{equation}{section}
\newcommand{\lyxaddress}[1]{
	\par {\raggedright #1
	\vspace{1.4em}
	\noindent\par}
}

\@ifundefined{date}{}{\date{}}
\makeatother

\begin{document}
\title{\textbf{A comparative study of quaternionic rotational Dirac equation
and its interpretation}}
\author{\textbf{B. C. Chanyal$^{\ast}$ and Sandhya$^{\dagger}$}}
\maketitle

\lyxaddress{\begin{center}
\textit{Department of Physics, G. B. Pant University of Agriculture
and Technology, Pantnagar-263145, Uttarakhand, India}\\
\textit{Email: }\textbf{$^{\ast}$}\textit{bcchanyal@gmail.com, bcchanyal@gbpuat.ac.in}\\
\textit{$^{\dagger}$karnatak18@gmail.com}
\par\end{center}}
\begin{abstract}
In this study, we develop the generalized Dirac like four-momentum
equation for rotating spin-half particles in four-dimensional quaternionic
algebra. The generalized quaternionic Dirac equation consists the
rotational energy and angular momentum of particle and anti-particle.
Accordingly, we also discuss the four vector form of quaternionic
relativistic mass, moment of inertia and rotational energy-momentum
in Euclidean space-time. The quaternionic four angular momentum (i.e.
the rotational analogy of four linear momentum) predicts the dual
energy (rest mass energy and pure rotational energy) and dual momentum
(linear like momentum and pure rotational momentum). Further, the
solutions of quaternionic rotational Dirac energy-momentum are obtained
by using one, two and four-component of quaternionic spinor. We also
demonstrate the solutions of quaternionic plane wave equation which
gives the rotational frequency and wave propagation vector of Dirac
particles and anti-particles in terms of quaternionic form.



\end{abstract}

\section{Introduction}

The results of classical theory have a large impact on the fully appropriate
description of matter but this theory does not explain the behavior
of subatomic particles. The information about the small-scale behavior
of particles created the idea of quantum mechanics. Generally, the
quantum physics deals with the description of the state of particles
associated with the wave function. The angular momentum is an observable
essential element in quantum mechanics, where the quantum particles
can involve the orbital angular momentum and intrinsic angular momentum.
Further, Erwin Schr�dinger developed a fundamental wave equation for
non-relativistic microscopic particles but this equation does not
applicable to particles moving with relativistic velocity. To combine
the special theory of relativity with quantum mechanics, there has
been developed the relativistic quantum mechanics. In the same way,
Klein-Gordon and Dirac independently investigated the relativistic
wave equations by combining special relativity with quantum mechanics.
These relativistic wave equations describe the various phenomenons
that occur in high energy physics and are invariant under Lorentz
transformations. Dirac \textbf{\cite{key-1}} discussed a relativistic
quantum wave equation by using Hamiltonian operator to overcome the
difficulties arising in Klein-Gordon equation. As we know that the
conservation of energy and angular momentum are one of the mandatory
conservation laws to check the validity of Dirac particles. Keeping
in mind the conservation laws of energy and angular momentum of a
rotating particle, in this paper, we proposed a quaternionic Dirac
equation that consists not only the energy representation but also
shows the angular momentum representation of spin-1/2 particles. The
quaternion number \cite{key-2} is basically an extension of complex
numbers. Although, there are four types of norm-division algebras,
i.e. real, complex, quaternion and octonion algebra. The division
algebra is defined as an algebra in which all non-zero elements have
their inverse under multiplication \cite{key-3}. Nowadays, the quaternionic
algebra is a popular algebra to study the various theories in modern
theoretical physics. The quaternionic algebra is associative and commutative
under addition but not commutative under multiplication. Thus, this
algebra form a group under multiplication but not an Abelian group,
is also called the division ring.

Many researchers have attempted the formulation of usual Dirac equation
for free particles in terms of quaternionic algebra. Firstly,\textbf{
}Rotelli \cite{key-4} developed the Dirac equation in quaternionic
four fields. The another version of quaternionic Dirac equation has
been studied by Rawat et. al \cite{key-5} with the description of
quaternionic spinors for positive and negative energy solutions. As
such, the quaternionic form of Dirac equation with the connection
of supersymmetric quantum mechanics has also been discussed \cite{key-6}.
Besides, the quaternionic algebra has also been used to describe the
rotational motion of a rigid body where the quaternionic unit elements
are directly connected with matrices of rotational group $SO(3)$
\cite{key-7}. Further, the quaternion algebra explained the special
theory of relativity \cite{key-8}, Dirac Lagrangian \cite{key-9},
superluminal transformations for tachyons \cite{key-10}, wave equation
in curved space-time \cite{key-11}, electromagnetic-field equations
\cite{key-12}, gravi-electromagnetism \cite{key-13}, quantum mechanics
\cite{key-14,key-15}, particle in a relativistic box \cite{key-16}
and dual magneto-hydrodynamics of dyonic cold plasma \cite{key-17}.
On the other hand, many authors \cite{key-18}-\cite{key-37} have
used hyper-complex algebras to study the several theories in different
branches of physics. Chanyal \cite{key-38,key-39} proposed a novel
idea on the quaternionic covariant theory of relativistic quantized
electromagnetic fields of dyons. Carmeli \cite{key-40,key-41,key-42}
discussed the various fundamental equations of quantum mechanics viz.
Klein-Gordon equation, Schr�dinger equation, Weyl equation\textbf{
}and the Dirac field equation of rotating particles on $R\times S^{3}$
topological space. Keeping in mind the properties of rotating spin
1/2 particles, we generalize the Carmeli's field theory in terms of
quaternionic field. The benefit of generalized quaternionic field
is that, we can analysis the four-momentum representation of a particle
in a single equation i.e. energy (as a scalar component) and momentum
(as a vector component). In present study, starting with quaternionic
algebra and its representation to $SU(2)$ group (i.e. an isomorphic
to orthogonal group $SO(3)$), we construct the generalized Dirac
like energy-momentum equation for rotating particles in four-dimensional
quaternionic space-time. We define the quaternionic moment of inertia
and rotational energy-momentum of a rotating particle by using four-relativistic
mass, four-spaces and four-momentum. A novel approach to unified quaternionic
Dirac like equation contains the rotational energy (corresponding
to the coefficient of quaternionic pure scalar unit element) and the
rotational momentum (corresponding to the coefficient of quaternionic
pure vector unit elements). Accordingly, the rotational energy and
momentum solutions are obtained by using one, two and four-component
spinor forms of quaternionic wave function. We have calculated the
solutions of rotational energy and rotational momentum for particles
with spin up and spin down states. To considering the wave nature
of spin half particles (or anti-particles), we have studied the general
form of quaternionic plane wave equation and developed the quaternionic
form of rotational frequency and wave propagation vector for particles
and anti-particles. This theory also point out the conservation law
of quaternionic four-momentum of particles.

\section{Quaternionic field representation}

The quaternionic field ($\mathbb{H}$) is expressed by a linear algebra
consist four unit elements known as quaternionic basis $\left(e_{0},\,e_{1},\,e_{2},\,e_{3}\right)$.
In a quaternionic field, the unit element $e_{0}$ is used to express
scalar field while the other unit elements $e_{1},\,e_{2},\,e_{3}$
are used to express vector fields. Thus, the quaternionic field algebra
can be expressed as
\begin{align}
\mathbb{H}\,\,= & \,\,e_{0}(\mathbb{H}^{\text{S}})+e_{j}(\mathbb{H}_{j}^{\text{V}})\,\,,\,\,\,\,\,\left(\forall\,\ensuremath{j=1,2,3}\right)\nonumber \\
\equiv & \,\,e_{0}h_{0}+e_{1}h_{1}+e_{2}h_{2}+e_{3}h_{3}\,,\label{eq:1}
\end{align}
where $\left(h_{0},\,h_{j}\right)$ are the real numbers corresponding
to quaternionic scalar-field $(\mathbb{H}^{\text{S}})$ and vector-field
$(\mathbb{H}^{\text{V}})$. If the real part of a quaternion is zero
then the quaternion have only vector field components called a pure
quaternion as $\mathbb{H}\rightarrow\mathbb{H}_{P}=\,\,\left(e_{1}h_{1}+e_{2}h_{2}+e_{3}h_{3}\right)$,
and if the imaginary part of a quaternion is zero then the quaternion
involve only scalar component called a real quaternion as $\mathbb{H\rightarrow H}_{R}=\,\,e_{0}h_{0}$.
The addition of two quaternions $\left(\mathbb{A},\,\mathbb{B}\right)\in\mathbb{H}$
produce a new quaternionic field as
\begin{align}
\mathbb{A}+\mathbb{B}\,\,= & \,\,\left(e_{0}a_{0}+e_{1}a_{1}+e_{2}a_{2}+e_{3}a_{3}\right)+\left(e_{0}b_{0}+e_{1}b_{1}+e_{2}b_{2}+e_{3}b_{3}\right)\nonumber \\
= & \,\,e_{0}\left(a_{0}+b_{0}\right)+e_{1}\left(a_{1}+b_{1}\right)+e_{2}\left(a_{2}+b_{2}\right)+e_{3}\left(a_{3}+b_{3}\right)\,.\label{eq:2}\\
\, & \,\,\,\,\,\,\,\,\,\,\,\,\,\,\,\,\,\,\,\,\,\,\,\,\,\,\,\,\,\,\,\,\,\,\,\,\,\,\,\,\,\,\,\,\,\,\,\,\,\,\,\,\,\,(a_{0},a_{j})\,\&\,(b_{0},b_{j})\in\mathbb{R}^{4}\nonumber 
\end{align}
The quaternions are associative as well as commutative under addition,
i.e.
\begin{align}
\mathbb{A}+\mathbb{B} & \,\,=\,\,\mathbb{B}+\mathbb{A}\,,\nonumber \\
\mathbb{A}+\left(\mathbb{B}+\mathbb{C}\right) & \,\,=\,\,\left(\mathbb{A}+\mathbb{B}\right)+\mathbb{C}\,,\,\,\,\,\,\,\,\forall\,\left(\mathbb{A},\,\mathbb{B},\,\mathbb{C}\right)\in\mathbb{H}\,.\label{eq:3}
\end{align}
The quaternionic algebra contains additive identity element zero as
\begin{align}
0= & \,\,e_{0}0+e_{1}0+e_{2}0+e_{3}0\,.\label{eq:4}
\end{align}
Every quaternion has its additive inverse which can be expressed in
the form of
\begin{align}
-\mathbb{H}= & \,\,e_{0}\left(-h_{0}\right)+e_{1}\left(-h_{1}\right)+e_{2}\left(-h_{2}\right)+e_{3}\left(-h_{3}\right)\,.\label{eq:5}
\end{align}
The quaternionic conjugate of given equation (\ref{eq:1}) can be
expressed as,
\begin{align}
\mathbb{\overline{H}}\,=\,\,e_{0}(\mathbb{H}^{\text{S}})-e_{j}(\mathbb{H}_{j}^{\text{V}})\,\equiv & \,\,e_{0}h_{0}-e_{1}h_{1}-e_{2}h_{2}-e_{3}h_{3}\,.\label{eq:6}
\end{align}
As such, the square of a quaternion is written as,
\begin{align}
H^{2}= & \mathbb{\,\,\overline{H}}\circ\mathbb{H}\,=\,\,\left(h_{0}^{2}+h_{1}^{2}+h_{2}^{2}+h_{3}^{2}\right)\,,\label{eq:7}
\end{align}
where $'\circ'$ indicated the quaternion multiplication. Here, the
quaternionic unit elements $(e_{0}\,,e_{1},\,e_{2},\,e_{3})$ are
followed the relations
\begin{align}
e_{0}^{2} & =\,1\,,\,\,e_{1}^{2}=e_{2}^{2}=e_{3}^{2}=\,-1\,,\nonumber \\
e_{0}e_{i} & =\,e_{i}e_{0}=e_{i}\,,\,\,e_{i}e_{j}=-\delta_{ij}e_{0}+\epsilon_{ijk}e_{k}\,,\,\,\,\forall\,(i,\,j,\,k=1,2,3)\label{eq:8}
\end{align}
where the symbol $\delta_{ij}$ is the Kronecker delta symbol having
value one for equal indies and zero for unequal indies, while $\epsilon_{ijk}$
is the Levi Civita three-index symbol taking value $\epsilon_{ijk}=+1$
for cyclic permutation, $\epsilon_{ijk}=-1$ for anti-cyclic permutation
and $\epsilon_{ijk}=0$ for any two repeated indices. The multiplication
of two quaternions can be expressed by using Table-1 as,
\begin{align}
\mathbb{A}\circ\mathbb{B}= & \,\,e_{0}\left(a_{0}b_{0}-a_{1}b_{1}-a_{2}b_{2}-a_{3}b_{3}\right)\nonumber \\
+ & e_{1}\left(a_{0}b_{1}+a_{1}b_{0}+a_{2}b_{3}-a_{3}b_{2}\right)\nonumber \\
+ & e_{2}\left(a_{0}b_{2}+a_{2}b_{0}+a_{3}b_{1}-a_{1}b_{3}\right)\nonumber \\
+ & e_{3}\left(a_{0}b_{3}+a_{3}b_{0}+a_{1}b_{2}-a_{2}b_{1}\right)\,,\label{eq:9}
\end{align}
which can be further simplify as
\begin{align}
\mathbb{A}\circ\mathbb{B}= & \,\,e_{0}\left(a_{0}b_{0}-\overrightarrow{a}\cdot\overrightarrow{b}\right)+e_{j}\left(a_{0}\overrightarrow{b}+b_{0}\overrightarrow{a}+\left(\overrightarrow{a}\times\overrightarrow{b}\right)_{j}\right)\,,\,\,\,\left(\forall\,\ensuremath{j=1,2,3}\right).\label{eq:10}
\end{align}
\begin{table}[t]
\centering{}\textbf{}%
\begin{tabular}{c|cccc}
\hline 
\textbf{$\circ$} & $e_{0}$ & \textbf{$e_{1}$} & \textbf{$e_{2}$} & \textbf{$e_{3}$}\tabularnewline
\hline 
\hline 
$e_{0}$ & $1$ & \textbf{$e_{1}$} & \textbf{$e_{2}$} & \textbf{$e_{3}$}\tabularnewline
\textbf{$e_{1}$} & \textbf{$e_{1}$} & \textbf{$-1$} & \textbf{$e_{3}$} & \textbf{$-e_{2}$}\tabularnewline
\textbf{$e_{2}$} & \textbf{$e_{2}$} & \textbf{$-e_{3}$} & \textbf{$-1$} & \textbf{$e_{1}$}\tabularnewline
\textbf{$e_{3}$} & \textbf{$e_{3}$} & \textbf{$e_{2}$} & \textbf{$-e_{1}$} & \textbf{$-1$}\tabularnewline
\hline 
\end{tabular}\caption{Quaternion multiplication table}
\end{table}

The quaternionic multiplication identity element can define as,
\begin{align}
1= & \,\,e_{0}1+e_{1}0+e_{2}0+e_{3}0\,.\label{eq:11}
\end{align}
Rather, the quaternionic field is associative but non-commutative
under multiplication operation, i.e.
\begin{align}
\left(\mathbb{A}\circ\mathbb{B}\right)\circ\mathbb{C} & =\,\,\mathbb{A}\circ\left(\mathbb{B}\circ\mathbb{C}\right)\,,\nonumber \\
\mathbb{A}\circ\mathbb{B} & \neq\,\,\mathbb{B}\circ\mathbb{A\,}.\label{eq:12}
\end{align}
Because the cross product of two vectors are always non-commutative
i.e. $\overrightarrow{a}\times\overrightarrow{b}=-\overrightarrow{b}\times\overrightarrow{a}$.
The two quaternions will be commutative only when the cross product
of their vectors are zero or the vectors $\overrightarrow{a}$ and
$\overrightarrow{b}$ are parallel to each other. Moreover, the norm
of a quaternion $\mathbb{A}$ becomes
\begin{align}
N\,\,= & \,\,\sqrt{\mathbb{\overline{A}\circ}\mathbb{A}}=\,\,\sqrt{a_{0}^{2}+a_{1}^{2}+a_{2}^{2}+a_{3}^{2}}\,.\label{eq:13}
\end{align}
The equation (\ref{eq:13}) also known the modulus of a quaternion
i.e. $|\mathbb{A}|$. Every non-zero element of a quaternion has an
inverse
\begin{align}
\mathbb{A}^{-1}= & \,\,\frac{\overline{\mathbb{A}}}{|\mathbb{A}|}\,.\label{eq:14}
\end{align}
Besides, a quaternion can also be represented in the form of split
quaternionic basis elements $u_{0},\,u_{0}^{\ast},\,u_{1},\,u_{1}^{\ast}$
as,
\begin{align}
\mathbb{A}=\,\, & u_{0}a+u_{0}^{\ast}a^{\ast}+u_{1}b+u_{1}^{\ast}b^{\ast}\,\,,\,\,\,\,\,\forall\,\left(a,\,a^{\ast},\,b,\,b^{\ast}\right)\in\mathbb{C}\nonumber \\
\equiv\,\, & u_{0}\left(h_{0}-ih_{3}\right)+u_{0}^{\ast}\left(h_{0}+ih_{3}\right)+u_{1}\left(h_{1}-ih_{2}\right)+u_{1}^{\ast}\left(h_{1}+ih_{2}\right)\,,\label{eq:15}
\end{align}
 where
\begin{align*}
u_{0}\,= & \,\,\frac{1}{2}\left(e_{0}+ie_{3}\right)\,\,,\text{\,\,\,\,\ensuremath{\,\,u_{0}^{\ast}\,=\,\,\frac{1}{2}\left(e_{0}-ie_{3}\right)}}\,,\\
u_{1}\,= & \,\,\frac{1}{2}\left(e_{1}+ie_{2}\right)\,\,,\,\,\,\,\,\,\text{\ensuremath{\,u_{1}^{\ast}\,=\,\,\frac{1}{2}\left(e_{1}-ie_{2}\right)} },
\end{align*}
 are written in the form of $2\times2$ matrix value realization as
\begin{align}
u_{0}= & \left(\begin{array}{cc}
1 & 0\\
0 & 0
\end{array}\right)\,\,,\,\,u_{0}^{\ast}=\left(\begin{array}{cc}
0 & 0\\
0 & 1
\end{array}\right)\,\,,\,\,u_{1}=\left(\begin{array}{cc}
0 & -i\\
0 & 0
\end{array}\right)\,\,,\,\,u_{1}^{\ast}=\left(\begin{array}{cc}
0 & 0\\
-i & 0
\end{array}\right)\,.\label{eq:16}
\end{align}
Thus, the quaternionic basis represents the following $2\times2$
matrix realization as
\begin{align}
e_{0}\mapsto & \left(\begin{array}{cc}
1 & 0\\
0 & 1
\end{array}\right),\,\,\,\,e_{1}\mapsto\left(\begin{array}{cc}
0 & -i\\
-i & 0
\end{array}\right),\,\,\,e_{2}\mapsto\left(\begin{array}{cc}
0 & -1\\
1 & 0
\end{array}\right),\,\,e_{3}\mapsto\left(\begin{array}{cc}
-i & 0\\
0 & i
\end{array}\right)\,\,,\label{eq:17}
\end{align}
where the determinant of each quaternionic basis gives unity i.e.,
$|e_{0}|=|e_{1}|=|e_{2}|=|e_{3}|=+1$. Interestingly, to considering
quaternion basis ($e_{0},\,e_{j}$), the algebra of ($e_{0},\,ie_{j}$)
is isomorphic to the algebra of tau-matrices ($\tau_{e_{0}},\,\tau_{e_{j}}$
for $j=1,2,3$) in $SU(2)$ group representation. Here, $\tau_{e_{0}}$
is $2\times2$ identity matrix corresponding to quaternion basis $e_{0}$
while $\tau_{e_{j}}$ are corresponding to pure quaternionic basis
called Pauli tau-matrices. In given equation (\ref{eq:15}), the $SU(2)$
representation is the set of all $2\times2$ complex matrices of determinant
positive one and satisfy $\mathbb{A}\mathbb{A}^{\dagger}=\mathbb{A}^{\dagger}\mathbb{A}=I_{2\times2}$,
i.e.
\begin{align}
\mathbb{A}\,\,=\,\, & \left(\begin{array}{cc}
h_{0}-ih_{3} & -(h_{2}+ih_{1})\\
h_{2}-ih_{1} & h_{0}+ih_{3}
\end{array}\right)\,\,\longmapsto\,\,\left(\begin{array}{cc}
a & -\left(ib\right)\\
\left(ib\right)^{\ast} & a^{\ast}
\end{array}\right)\,,\label{eq:18}
\end{align}
where $\mid a\mid^{2}+\mid b\mid^{2}=1$. For quaternionic rotational
group, if $\mathbb{A}\in SU(2)$ maps onto $R(\mathbb{A})\in SO(3)$,
then we may write
\begin{align}
R(\mathbb{A})_{j,k}\,\,=\,\, & \frac{1}{2}\text{Tr}(\tau_{e_{j}}\mathbb{A}\tau_{e_{k}}\mathbb{A}^{-1})\,,\,\,\,\,\,\,(\forall\,j,k=1,2,3).\label{eq:19}
\end{align}
As such, the simplest representations of the quaternionic basis can
also be expressed by the multiplication of $(-i)$ with 2 \texttimes{}
2 Pauli tau-matrices as,
\begin{align}
e_{0}= & \,\,\left(\begin{array}{cc}
1 & 0\\
0 & 1
\end{array}\right)\longmapsto\tau_{e_{0}}\,,\,\,\,\,\,\,\,\,\,\,\,\,\,\,\,\,\,e_{1}=-i\left(\begin{array}{cc}
0 & 1\\
1 & 0
\end{array}\right)\longmapsto-i\tau_{e_{1}}\,,\nonumber \\
e_{2}= & -i\left(\begin{array}{cc}
0 & -i\\
i & 0
\end{array}\right)\longmapsto-i\tau_{e_{2}}\,,\,\,\,\,\,e_{3}=-i\left(\begin{array}{cc}
1 & 0\\
0 & -1
\end{array}\right)\longmapsto-i\tau_{e_{3}}\,,\label{eq:20}
\end{align}
where $e_{j}\cong(-i\tau_{e_{j}})\,\,j=1,2,3$ are also defined a
realization of infinitesimal rotations of three dimensions pure quaternionic
space. Moreover, the quaternions can be use to study the rotational
motion of a spin-1/2 particles because it is identical to rotational
$\tau-$matrices\textbf{ }\cite{key-43}. We summarize the properties
of quaternionic basis and tau-matrices in Table-2.
\begin{table}[t]
\begin{onehalfspace}
\begin{centering}
\begin{tabular}{c|c|c}
\hline 
\textbf{Properties} & \textbf{Quaternion basis} & \textbf{Tau-matrices}\tabularnewline
\hline 
\hline 
Square & $e_{0}^{2}=1$, $e_{j}^{2}=-1$ & $\left(\tau_{e_{0}}\right)^{2}=1$,~$(-i\tau_{e_{j}})^{2}=-1$\tabularnewline
Determinant & $1$ & $1$\tabularnewline
Eigen values & $\pm i$ & $\pm i$\tabularnewline
Trace & $0$ & $0$\tabularnewline
Multiplication & $e_{i}e_{j}=-\delta_{ij}e_{0}+\epsilon_{ijk}e_{k}$ & $\tau_{e_{i}}\tau_{e_{j}}=\delta_{ij}\tau_{e_{0}}+i\epsilon_{ijk}\tau_{e_{k}}$\tabularnewline
Commutation & $\left[e_{i},\,e_{j}\right]=2e_{k}\epsilon_{ijk}$ & $\left[\tau_{e_{i}},\,\tau_{e_{j}}\right]=2i\epsilon_{ijk}\tau_{e_{k}}$\tabularnewline
Anti-commutation & $\left\{ e_{i},\,e_{j}\right\} =-2e_{0}\delta_{ij}$ & $\left\{ \tau_{e_{i}},\,\tau_{e_{j}}\right\} =2\delta_{ij}\tau_{e_{0}}$\tabularnewline
\hline 
\end{tabular}
\par\end{centering}
\end{onehalfspace}
\caption{Properties of quaternionic basis and $\tau-$matrices}
\end{table}

\section{Quaternionic rotational energy-momentum}

Let us start with the quaternionic relativistic mass of Dirac particles
(e.g. electrons) that can be expressed in terms of four-masses\textbf{
}\cite{key-44},
\begin{align}
\mathbb{M}\,= & \,\,\,\,e_{0}\frac{\mathcal{E}_{0}}{c^{2}}+e_{1}\mid\frac{p_{1}}{v_{1}}\mid+e_{2}\mid\frac{p_{2}}{v_{2}}\mid+e_{3}\mid\frac{p_{3}}{v_{3}}\mid\nonumber \\
\simeq & \,\,e_{0}m_{0}+\sum_{j=1}^{3}e_{j}m_{j}\,\,,\label{eq:21}
\end{align}
where $m_{0}\sim\mathcal{E}_{0}/c^{2}$ indicates the rest mass and
$m_{1}$, $m_{2}$, $m_{3}$ are the moving masses of particle having
velocities $v_{1},\,v_{2}$, $v_{3}$ corresponding to quaternionic
basis $e_{1},\,e_{2}$, $e_{3},$ respectively. It should be notice
that the quaternionic scalar field associated with the coefficient
of $e_{0}$ while the quaternionic vector field associated with the
coefficient of $e_{j}$. If vector part is zero then the real quaternion
expresses only the rest mass ($m_{0}$), and if rest mass-energy of
a particle is zero then the pure quaternion experiences the motion
of particle. As such, the quaternionic space-time position $(\mathbb{R})$
can also be expressed as
\begin{align}
\mathbb{R}\,= & \:e_{0}r_{0}+e_{1}r_{1}+e_{2}r_{2}+e_{3}r_{3}\,\,,\label{eq:22}
\end{align}
where $r_{0}$ is the scalar component considering as time, while
$r_{1},\,r_{2},\,r_{3}$ are the three spacial components in Euclidean
space-time. Now, the moment of inertia in quaternionic form becomes,
\begin{align}
\mathbb{I}= & \,\,\mathbb{M}\left(\mathbb{\overline{R}}\circ\mathbb{R}\right)=\mathbb{M}\left(r_{0}^{2}+r_{1}^{2}+r_{2}^{2}+r_{3}^{2}\right)\nonumber \\
= & \,\,\mathbb{M}R^{2}\,\,.\label{eq:23}
\end{align}
Thus, we have
\begin{align}
\mathbb{I}= & \,\,e_{0}I_{0}+e_{1}I_{1}+e_{2}I_{2}+e_{3}I_{3}\,\,,\label{eq:24}
\end{align}
where $I_{0}=m_{0}R^{2}$ is considered the moment of inertia along
scalar axis $e_{0}$ while $I_{1}=m_{1}R^{2}$, $I_{1}=m_{2}R^{2}$
and $I_{3}=m_{3}R^{2}$ are the moment of inertia along pure-quaternionic
axes $e_{1},\,e_{2},\,e_{3},$ respectively. Keeping in mind the rotational
analog of translation motion, we can define the quaternionic four-angular
momentum as,
\begin{align}
\mathbb{L=} & \,\,\mathbb{R}\circ\mathbb{P}\,\,,\label{eq:25}
\end{align}
where $\mathbb{P}\mapsto\{e_{0}p_{0},\,e_{1}p_{1},\,e_{2}p_{2},\,e_{3}p_{3}\}$
is the quaternionic linear four-momentum. Thus, from equation (\ref{eq:25}),
we get
\begin{align}
\mathbb{L}= & \,\,e_{0}\left(r_{0}m_{0}c-r_{1}m_{1}v_{1}-r_{2}m_{2}v_{2}-r_{3}m_{3}v_{3}\right)\nonumber \\
+ & e_{1}\left(r_{0}m_{1}v_{1}+r_{1}m_{0}c+r_{2}m_{3}v_{3}-r_{3}m_{2}v_{2}\right)\nonumber \\
+ & e_{2}\left(r_{0}m_{2}v_{2}+r_{2}m_{0}c+r_{3}m_{1}v_{1}-r_{1}m_{3}v_{3}\right)\nonumber \\
+ & e_{3}\left(r_{0}m_{3}v_{3}+r_{3}m_{0}c+r_{1}m_{2}v_{2}-r_{2}m_{1}v_{1}\right)\,\,.\label{eq:26}
\end{align}
It can be reduces as,
\begin{align}
\mathbb{L}= & \,\,e_{0}L_{0}+e_{1}L_{1}+e_{2}L_{2}+e_{3}L_{3}\nonumber \\
= & \,\,e_{0}\left(L_{00}-L_{11}-L_{22}-L_{33}\right)+e_{1}\left(L_{01}+L_{10}+L_{23}-L_{32}\right)\nonumber \\
\, & +e_{2}\left(L_{02}+L_{20}+L_{31}-L_{13}\right)+e_{3}\left(L_{03}+L_{30}+L_{12}-L_{12}\right)\,,\label{eq:27}
\end{align}
where $L_{0}$ is purely a scalar component can be indicated by quaternionic
rotational energy $\left(E_{0}\right)$, while the vector components
($L_{j},\,j=1,2,3$) indicated the quaternionic rotational angular-momentum.
Now, the quaternionic rotational energy $E_{0}\sim L_{0}$ can be
expressed by
\begin{align}
E_{0}= & \,\,r_{0}p_{0}-\left(\overrightarrow{r}\cdot\overrightarrow{p}\right)\,\,,\,\,\,\,\,\,(\text{coefficient of }e_{0})\,.\label{eq:28}
\end{align}
Here, in quaternionic formalism the quaternionic rotational energy
consists a couple of energy. We may consider the first scalar term
($r_{0}p_{0}$) represents the rest mass-energy while the second scalar
term $\left(\overrightarrow{r}\cdot\overrightarrow{p}\right)$ represents
the moving (projectional) energy. In the same way, the quaternionic
angular momentum can then be written as,
\begin{align}
L_{j}= & \,\,\left(r_{0}\overrightarrow{p}+p_{0}\overrightarrow{r}\right)+\left(\overrightarrow{r}\times\overrightarrow{p}\right)_{j}\,\,,\,\,\,\,\,\,(\text{coefficient of }e_{j})\,\,.\label{eq:29}
\end{align}
In equation (\ref{eq:29}), the first term indicates the longitudinal
component (irrotational momentum) while the second term indicates
the transverse component (rotational momentum) of quaternionic four-momentum.
Therefore, we may describe the quaternionic form of rotational energy-momentum
as following specific cases:\smallskip{}

\textbf{Case-1:} Conditionally, for the pure quaternionic field the
scalar coefficient is taken zero i.e., $r_{0}=p_{0}=0.$ Then, we
get the usual three dimensional form of rotational energy and angular
momentum in vector field, i.e.

\begin{align}
\mid E_{0}\mid\,\simeq & \,\,\left(\overrightarrow{r}\cdot\overrightarrow{p}\right)\,\,,\,\,\,\,\,\,\,\,\text{(pure rotational energy)}\label{eq:30}\\
\overrightarrow{L}\,\simeq & \,\,\left(\overrightarrow{r}\times\overrightarrow{p}\right)\,\,,\,\,\,\,\,\text{(pure angular momentum)}\,.\label{eq:31}
\end{align}

\textbf{Case-2:} If pure quaternion part is zero i.e. $\overrightarrow{r}=\overrightarrow{p}=0$,
then we get the pure scalar field as
\begin{align}
E_{0}\,\,\simeq & \,\,r_{0}p_{0}\,,\,\,\,\overrightarrow{L}\,\,\simeq\,\,0\,\,,\label{eq:32}
\end{align}
which shows that there will be only rest mass-energy having no rotating
motion.\smallskip{}

\textbf{Case-3:} If the quaternionic variables $\mathbb{R}$ and $\mathbb{P}$
are mutually interchange as $\mathbb{R}\rightarrow\mathbb{P}$ and
$\mathbb{P}\rightarrow\mathbb{R}$, then the rotational energy $(E_{0})$
remains unaffected but the angular momentum $(\overrightarrow{L})$
become changed, i.e.,
\begin{align}
E_{0}\left(\mathbb{R}\Leftrightarrow\mathbb{P}\right)= & \,\,r_{0}p_{0}-\left(\overrightarrow{r}\cdot\overrightarrow{p}\right)\,\,,\nonumber \\
\overrightarrow{L}\left(\mathbb{R}\Leftrightarrow\mathbb{P}\right)= & \,\,\left(p_{0}\overrightarrow{r}+r_{0}\overrightarrow{p}\right)-\left(\overrightarrow{r}\times\overrightarrow{p}\right)\,\,.\label{eq:33}
\end{align}

\section{Quaternionic Rotational Dirac (QRD) equation}

Let us start with the Dirac equation for free particle as
\begin{align}
\left(\overrightarrow{\alpha}\cdot\overrightarrow{p}-\mathscr{\beta}mc^{2}\right)\psi & =\,\,0\,\,,\label{eq:34}
\end{align}
where $\overrightarrow{\alpha}$ and $\beta$ are the Dirac matrices.
Now, to check the energy and momentum relations of an electron rotating
in quaternionic space-time, we can extend the Dirac equation (\ref{eq:34})
in term of QRD equation form, i.e.
\begin{align}
\left(\underline{\mathbb{A}}\circ\mathbb{L}-\mathscr{B}\lambda^{2}\mathbb{I}\right)\circ\boldsymbol{\Psi} & =\,\,0\,\,,\label{eq:35}
\end{align}
where $\underline{\mathbb{A}}$, $\mathbb{L}$, $\mathscr{B}$, $\mathbb{I}$
and $\boldsymbol{\Psi}$ are quaternionic variables considering for
the rotational analogy of $\overrightarrow{\alpha}$, $\overrightarrow{p}$,
$\beta$, $m$ and $\psi$, respectively. For rotating particles the
speed of light $c$ can be replaced by the maximum speed $\lambda$
\cite{key-42} as $\lambda=\,c\left(\frac{\mathbb{M}}{\mathbb{I}}\right)^{\frac{1}{2}}=\,\frac{c}{R}\,$
for rotating particle, where $R=\sqrt{r_{0}^{2}+r_{1}^{2}+r_{2}^{2}+r_{3}^{2}}$.
Using irreducible representation of $SU(2)$ group, we may write the
rotational Dirac matrices \textbf{$\underline{\mathbb{A}}$ }with
quaternionic structure as
\begin{align}
\underline{\mathbb{A}}= & \,\,\left(D^{0}\left(A\right),\,\,D^{j}\left(A\right)\right)\,\,,\,\,\,\,\,\left(\forall\,j=1,2,3\right)\,\,,\label{eq:36}
\end{align}
where the matrices $D^{0}\left(A\right)$ and $D^{j}\left(A\right)$
are quaternionic $D-$matrices, can be define as

\begin{align}
D^{0}\left(A\right)= & \left(\begin{array}{cc}
\tau_{e_{0}} & 0\\
0 & \tau_{e_{0}}
\end{array}\right)\,\,,\,\,D^{j}\left(A\right)=\left(\begin{array}{cc}
0 & \tau_{e_{j}}\\
\tau_{e_{j}} & 0
\end{array}\right)\,,\,\,\,\,\left(\forall\,j=1,2,3\right)\,,\label{eq:37}
\end{align}
and the $\mathscr{B}-$matrix is
\begin{align}
\mathscr{B}= & \left(\begin{array}{cc}
1 & 0\\
0 & -1
\end{array}\right)\,.\label{eq:38}
\end{align}
Now, using quaternion multiplication the first term $\underline{\mathbb{A}}\circ\mathbb{L}$
of QRD equation (\ref{eq:35}) can be expressed by
\begin{align}
\underline{\mathbb{A}}\circ\mathbb{L}\,= & \,\,e_{0}\left[D^{0}\left(A\right)E_{0}-\lambda D^{1}\left(A\right)L_{1}-\lambda D^{2}\left(A\right)L_{2}-\lambda D^{3}\left(A\right)L_{3}\right]\nonumber \\
+ & e_{1}\left[\lambda D^{0}\left(A\right)L_{1}+D^{1}\left(A\right)E_{0}+\lambda D^{2}\left(A\right)L_{3}-\lambda D^{3}\left(A\right)L_{2}\right]\nonumber \\
+ & e_{2}\left[\lambda D^{0}\left(A\right)L_{2}+D^{2}\left(A\right)E_{0}+\lambda D^{3}\left(A\right)L_{1}-\lambda D^{1}\left(A\right)L_{3}\right]\nonumber \\
+ & e_{3}\left[\lambda D^{0}\left(A\right)L_{3}+D^{3}\left(A\right)E_{0}+\lambda D^{1}\left(A\right)L_{2}-\lambda D^{2}\left(A\right)L_{1}\right]\,\,,\label{eq:39}
\end{align}
which can further reduces as
\begin{align}
\underline{\mathbb{A}}\circ\mathbb{L}\,= & \,\,e_{0}\left[D^{0}\left(A\right)E_{0}-\lambda\left(\overrightarrow{D}\left(A\right)\cdot\overrightarrow{L}\right)\right]\nonumber \\
+ & e_{j}\left[\lambda D^{0}\left(A\right)L_{j}+D^{j}\left(A\right)E_{0}+\lambda\left(\overrightarrow{D}\left(A\right)\times\overrightarrow{L}\right)_{j}\right]\,\,,\,\,\,\,\forall\,\left(j=1,2,3\right)\,.\label{eq:40}
\end{align}
Correspondingly, the second term of QRD equation (\ref{eq:35}) i.e.
$\mathscr{B}\lambda^{2}\mathbb{I}$ can be written as,
\begin{align}
\mathscr{B}\lambda^{2}\mathbb{I}= & \,\,e_{0}\mathscr{B}\lambda^{2}I_{0}+e_{1}\mathscr{B}\lambda^{2}I_{1}+e_{2}\mathscr{B}\lambda^{2}I_{2}+e_{3}\mathscr{B}\lambda^{2}I_{3}\,\,.\label{eq:41}
\end{align}
Therefore, from equations (\ref{eq:40}) and (\ref{eq:41}) the generalized
QRD equation can be expressed as
\begin{align}
[e_{0}\{D^{0}\left(A\right)E_{0}-\lambda\left(\overrightarrow{D}\left(A\right)\cdot\overrightarrow{L}\right) & -\mathscr{B}\lambda^{2}I_{0}\}\nonumber \\
+\,e_{j}\{\lambda D^{0}\left(A\right)L_{j}+D^{j}\left(A\right)E_{0} & +\lambda\left(\overrightarrow{D}\left(A\right)\times\overrightarrow{L}\right)_{j}-\mathscr{B}\lambda^{2}I_{j}\}]\circ\boldsymbol{\Psi}=0\,.\label{eq:42}
\end{align}
Interestingly, the QRD equation (\ref{eq:42}) consist both scalar
and vector components that gives not only the rotational energy but
also gives the angular momentum of electrons.
\begin{align*}
\text{Real quaternionic field }(\text{corresponding to }e_{0})\,\,\,:\longmapsto & \,\,\text{\,\,Dirac rotational energy}\,,\\
\text{Pure quaternionic field }(\text{corresponding to }e_{j})\,\,\,:\longmapsto & \,\,\text{\,\,Dirac rotational momentum\,.}
\end{align*}
To explain the dual nature of quaternionic unified rotational energy-momentum
solutions, we start with quaternionic spinor ($\boldsymbol{\Psi}$)
with scalar and vector fields as \cite{key-6},
\begin{align}
\boldsymbol{\Psi}= & \,\,e_{0}\Psi_{0}+e_{1}\Psi_{1}+e_{2}\Psi_{2}+e_{3}\Psi_{3}\nonumber \\
= & \left(\Psi_{0}+e_{1}\Psi_{1}\right)+e_{2}\left(\Psi_{2}-e_{1}\Psi_{3}\right)\nonumber \\
= & \,\,\Psi_{a}+e_{2}\Psi_{b},\label{eq:43}
\end{align}
where $\Psi_{a}=\left(\Psi_{0}+e_{1}\Psi_{1}\right)$ and $\Psi_{b}=\left(\Psi_{2}-e_{1}\Psi_{3}\right)$.
Further, for two and four components form, the quaternionic spinors
can be written as
\begin{align}
\boldsymbol{\Psi}= & \left(\begin{array}{c}
\Psi_{a}\\
\Psi_{b}
\end{array}\right)\,,\,\,\,\,\,\,\text{(two component form),}\label{eq:44}
\end{align}
and
\begin{align}
\boldsymbol{\Psi}= & \left(\begin{array}{c}
\Psi_{0}\\
\Psi_{1}\\
\Psi_{2}\\
\Psi_{3}
\end{array}\right)\,,\,\,\,\,\,\,\,\text{(four component form)}.\label{eq:45}
\end{align}
Now, in next section we shall use one, two and four-component spinors
to determine the solutions of quaternionic rotational energy and angular
momentum.
\begin{onehalfspace}

\section{The energy solutions of QRD equation}
\end{onehalfspace}

In order to attempt the energy solutions of QRD equation, we equate
the scalar components (coefficient of $e_{0}$) in given equation
(\ref{eq:42}) as
\begin{align}
\left[D^{0}\left(A\right)E_{0}-\lambda\left(\overrightarrow{D}\left(A\right)\cdot\overrightarrow{L}\right)-\mathscr{B}\lambda^{2}I_{0}\right]\boldsymbol{\Psi}= & \,\,0\,\,,\label{eq:46}
\end{align}
substituting the value of $D^{0}\left(A\right),$ $\overrightarrow{D}\left(A\right)$
and $\mathscr{B}$ from equations (\ref{eq:37}) and (\ref{eq:38}),
and obtain the following matrix form
\begin{align}
\left(\begin{array}{cc}
E_{0}-\lambda^{2}I_{0} & -i\lambda\left(\overrightarrow{e}\cdot\overrightarrow{L}\right)\\
-i\lambda\left(\overrightarrow{e}\cdot\overrightarrow{L}\right) & E_{0}+\lambda^{2}I_{0}
\end{array}\right)\left(\begin{array}{c}
\Psi_{a}\\
\Psi_{b}
\end{array}\right)= & \,\,0\,\,,\label{eq:47}
\end{align}
which gives
\begin{align}
\left(E_{0}-\lambda^{2}I_{0}\right)\Psi_{a}-i\lambda\left(\overrightarrow{e}\cdot\overrightarrow{L}\right)\Psi_{b}= & \,\,0\,\,,\label{eq:48}\\
\left(E_{0}+\lambda^{2}I_{0}\right)\Psi_{b}-i\lambda\left(\overrightarrow{e}\cdot\overrightarrow{L}\right)\Psi_{a}= & \,\,0\,\,,\label{eq:49}
\end{align}
where $\overrightarrow{e}\rightarrow(e_{1},\,e_{2},\,e_{3})$. Equations
(\ref{eq:48}) and (\ref{eq:49}) are represented the positive and
negative energies of a rotating Dirac particle, respectively. The
values of $\Psi_{a}$ and $\Psi_{b}$ are coupled with the function
of rotational energy and angular momentum ($E_{0},\overrightarrow{L}$),
i.e.
\begin{align}
\Psi_{0}(E_{0},\overrightarrow{L})= & \,\,\frac{\lambda}{E_{0}-\lambda^{2}I_{0}}i\left(\overrightarrow{e}\cdot\overrightarrow{L}\right)\Psi_{2}(E_{0},\overrightarrow{L})\,\,,\label{eq:50}\\
\Psi_{1}(E_{0},\overrightarrow{L})= & \,\,\frac{\lambda}{E_{0}-\lambda^{2}I_{0}}i\left(\overrightarrow{e}\cdot\overrightarrow{L}\right)\Psi_{3}(E_{0},\overrightarrow{L})\,\,,\label{eq:51}\\
\Psi_{2}(E_{0},\overrightarrow{L})= & \,\,\frac{\lambda}{E_{0}+\lambda^{2}I_{0}}i\left(\overrightarrow{e}\cdot\overrightarrow{L}\right)\Psi_{0}(E_{0},\overrightarrow{L})\,\,,\label{eq:52}\\
\Psi_{3}(E_{0},\overrightarrow{L})= & \,\,\frac{\lambda}{E_{0}+\lambda^{2}I_{0}}i\left(\overrightarrow{e}\cdot\overrightarrow{L}\right)\Psi_{1}(E_{0},\overrightarrow{L})\,\,,\label{eq:53}
\end{align}
which yields,
\begin{align}
\Psi_{\varLambda}(E_{0},\overrightarrow{L})-i\Omega_{+}(E_{0})\left[\overrightarrow{e}\cdot\overrightarrow{L}\right]\Psi_{\varLambda+2}(E_{0},\overrightarrow{L})= & 0\,,\,\left(\varLambda=0,1\right)\,(\text{for\,positive\,rotational\,energy})\nonumber \\
\Psi_{\varLambda}(E_{0},\overrightarrow{L})-i\Omega_{-}(E_{0})\left[\overrightarrow{e}\cdot\overrightarrow{L}\right]\Psi_{\varLambda-2}(E_{0},\overrightarrow{L})= & 0\,,\,\left(\varLambda=2,3\right)\,(\text{for\,negative\,rotational\,energy}).\label{eq:54}
\end{align}
Equation (\ref{eq:54}) shows the complex behavior of quaternionic
quantum wave function associated with the interaction between quaternion
spin and orbital angular momentum $\left(\overrightarrow{e}\cdot\overrightarrow{L}\right)$.
It should be notice that the imaginary unit $i$ placed with quaternionic
basis to represent the rotational matrices, so that, $ie_{j}\cong\tau_{e_{j}}\,,\,j=1,2,3$.
Here, $\Omega_{\pm}(E_{0})=\frac{\lambda}{E_{0}\mp\lambda^{2}I_{0}}$
is a constant used for a quaternionic rotational energy (i.e. $\Omega_{+}(E_{0})$
corresponding to positive energy and $\Omega_{-}(E_{0})$ corresponding
to negative energy).

\subsection{One component solutions}

In this case, we consider $\Psi_{0}=1,\text{ \ensuremath{\Psi_{1}=0}}$
for spin up and $\Psi_{0}=0$, $\Psi_{1}=1$ for spin down positive
energy solutions. Then, we obtain
\begin{align}
\boldsymbol{\Psi}\rightarrow\boldsymbol{\Psi}^{\uparrow+}(E_{0},\overrightarrow{L})= & \,\,N_{+}^{E}\left(1+e_{2}\frac{i\lambda\left(\overrightarrow{e}\cdot\overrightarrow{L}\right)}{E_{0}+\lambda^{2}I_{0}}\right)\,\,,\label{eq:55}\\
\boldsymbol{\Psi}\rightarrow\boldsymbol{\Psi}^{\downarrow+}(E_{0},\overrightarrow{L})= & \,\,N_{+}^{E}e_{1}\left(1+e_{2}\frac{i\lambda\left(\overrightarrow{e}\cdot\overrightarrow{L}\right)}{E_{0}+\lambda^{2}I_{0}}\right)\,,\label{eq:56}
\end{align}
where the normalization constant $N_{+}^{E}=\,\,\frac{E_{0}+\lambda^{2}I_{0}}{\sqrt{\left(E_{0}+\lambda^{2}I_{0}\right)^{2}+\lambda^{2}L_{j}^{2}}}$.
Similarly, for negative energy solutions with spin up and spin down
states, we get
\begin{align}
\boldsymbol{\Psi}\rightarrow\boldsymbol{\Psi}^{\uparrow-}(E_{0},\overrightarrow{L})= & \,\,N_{-}^{E}\left(\frac{i\lambda\left(\overrightarrow{e}\cdot\overrightarrow{L}\right)}{E_{0}-\lambda^{2}I_{0}}+e_{2}\right)\,,\label{eq:57}
\end{align}
\begin{alignat}{1}
\boldsymbol{\Psi}\rightarrow\boldsymbol{\Psi}^{\downarrow-}(E_{0},\overrightarrow{L})= & \,\,N_{-}^{E}e_{1}\left(\frac{i\lambda\left(\overrightarrow{e}\cdot\overrightarrow{L}\right)}{E_{0}-\lambda^{2}I_{0}}+e_{2}\right)\,\,.\label{eq:58}
\end{alignat}
where $N_{-}^{E}=\,\,\frac{E_{0}-\lambda^{2}I_{0}}{\sqrt{\left(E_{0}-\lambda^{2}I_{0}\right)^{2}+\lambda^{2}L_{j}^{2}}}$.
It should be notice that in one component formalism, all positive
and negative energy (spin up and spin down) spinors associated with
quaternionic basis with the fields corresponding to particle and antiparticle.

\subsection{Two component solutions}

The two component solutions corresponding to positive and negative
energy with spin up state are expressed in quaternionic from as,
\begin{align}
\boldsymbol{\Psi}\rightarrow\boldsymbol{\Psi}^{\uparrow+}(E_{0},\overrightarrow{L})= & \,\,N_{+}^{E}\left(\begin{array}{c}
1\\
\frac{i\lambda\left(\overrightarrow{e}\cdot\overrightarrow{L}\right)}{E_{0}+\lambda^{2}I_{0}}
\end{array}\right)\,\label{eq:59}\\
\boldsymbol{\Psi}\rightarrow\boldsymbol{\Psi}^{\uparrow-}(E_{0},\overrightarrow{L})= & \,\,N_{-}^{E}\left(\begin{array}{c}
\frac{i\lambda\left(\overrightarrow{e}\cdot\overrightarrow{L}\right)}{E_{0}-\lambda^{2}I_{0}}\\
1
\end{array}\right)\,\,\label{eq:60}
\end{align}
Similarly, for positive and negative energy with spin down states,
\begin{align}
\boldsymbol{\Psi}\rightarrow\boldsymbol{\Psi}^{\downarrow+}(E_{0},\overrightarrow{L})= & \,\,N_{+}^{E}e_{1}\left(\begin{array}{c}
1\\
\frac{-i\lambda\left(\overrightarrow{e}\cdot\overrightarrow{L}\right)}{E_{0}+\lambda^{2}I_{0}}
\end{array}\right)\,\,\simeq\,\,-iN_{+}^{E}\left(\begin{array}{c}
\frac{-i\lambda\left(\overrightarrow{e}\cdot\overrightarrow{L}\right)}{E_{0}+\lambda^{2}I_{0}}\\
1
\end{array}\right)\,\,.\label{eq:61}\\
\boldsymbol{\Psi}\rightarrow\boldsymbol{\Psi}^{\downarrow-}(E_{0},\overrightarrow{L})= & \,\,-N_{-}^{E}e_{1}\left(\begin{array}{c}
\frac{-i\lambda\left(\overrightarrow{e}\cdot\overrightarrow{L}\right)}{E_{0}-\lambda^{2}I_{0}}\\
1
\end{array}\right)\,\,\simeq\,\,iN_{-}^{E}\left(\begin{array}{c}
1\\
\frac{-i\lambda\left(\overrightarrow{e}\cdot\overrightarrow{L}\right)}{E_{0}-\lambda^{2}I_{0}}
\end{array}\right)\label{eq:62}
\end{align}
Notice that, the quaternionic basis element $e_{3}$ shows the diagonal
matrix which does not able to change the state of orientation of spin-1/2
particles, while the basis $e_{1}\text{and}e_{2}$ show off-diagonal
matrices whose can transform the state of orientation of spin-1/2
particles. The quaternionic unified of two component solution can
now be written as,
\begin{align}
\boldsymbol{\Psi}_{\text{Unified}}(E_{0},\overrightarrow{L})\,\,\simeq & \,\,N_{+}^{E}\left\{ \left(\begin{array}{c}
1\\
\frac{i\lambda\left(\overrightarrow{e}\cdot\overrightarrow{L}\right)}{E_{0}+\lambda^{2}I_{0}}
\end{array}\right)-i\left(\begin{array}{c}
\frac{-i\lambda\left(\overrightarrow{e}\cdot\overrightarrow{L}\right)}{E_{0}+\lambda^{2}I_{0}}\\
1
\end{array}\right)\right\} \nonumber \\
 & +N_{-}^{E}\left\{ \left(\begin{array}{c}
\frac{i\lambda\left(\overrightarrow{e}\cdot\overrightarrow{L}\right)}{E_{0}-\lambda^{2}I_{0}}\\
1
\end{array}\right)+i\left(\begin{array}{c}
1\\
\frac{-i\lambda\left(\overrightarrow{e}\cdot\overrightarrow{L}\right)}{E_{0}-\lambda^{2}I_{0}}
\end{array}\right)\right\} \,\,.\label{eq:63}
\end{align}
The unified spinor-function of two component solution shows a complex
behavior of quaternionic field where the each component predicts the
energy solution for particles (or anti particles) with their spinor.

\subsection{Four component solutions}

Like one and two component solutions, we can extent it into four component
solutions. In this case, we obtain the quaternionic four component
solutions for positive energy with spin up and down states as
\begin{align}
\boldsymbol{\Psi}\rightarrow\boldsymbol{\Psi}^{\uparrow+}(E_{0},\overrightarrow{L})\simeq\,\,N_{+}^{E} & \left(\begin{array}{c}
1\\
0\\
\frac{i\lambda\left(\overrightarrow{e}\cdot\overrightarrow{L}\right)}{E_{0}+\lambda^{2}I_{0}}\\
0
\end{array}\right)\,,\,\boldsymbol{\Psi}\rightarrow\boldsymbol{\Psi}^{\downarrow+}(E_{0},\overrightarrow{L})\simeq\,\,N_{+}^{E}\left(\begin{array}{c}
0\\
1\\
0\\
\frac{-i\lambda\left(\overrightarrow{e}\cdot\overrightarrow{L}\right)}{E_{0}+\lambda^{2}I_{0}}
\end{array}\right)\,,\label{eq:64}
\end{align}
Similarly, the negative energy solutions for spin up and down states
are
\begin{align}
\boldsymbol{\Psi}\rightarrow\boldsymbol{\Psi}^{\uparrow-}(E_{0},\overrightarrow{L})\simeq\,\,N_{-}^{E} & \left(\begin{array}{c}
\frac{i\lambda\left(\overrightarrow{e}\cdot\overrightarrow{L}\right)}{E_{0}-\lambda^{2}I_{0}}\\
0\\
1\\
0
\end{array}\right)\,,\,\boldsymbol{\Psi}\rightarrow\boldsymbol{\Psi}^{\downarrow-}(E_{0},\overrightarrow{L})\simeq\,\,N_{-}^{E}\left(\begin{array}{c}
0\\
\frac{-i\lambda\left(\overrightarrow{e}\cdot\overrightarrow{L}\right)}{E_{0}-\lambda^{2}I_{0}}\\
0\\
1
\end{array}\right)\,.\label{eq:65}
\end{align}
Interestingly, the one, two and four component solutions are isomorphic
to each other. In quaternionic formalism, the scalar term $\left(\overrightarrow{e}\cdot\overrightarrow{L}\right)$
can be used as the rotational helicity of particle, which represents
the quaternionic spin-orbit interaction energy. In right hand rotational
helicity the direction of spin is along to the direction of quaternionic
angular momentum, while for left hand rotational helicity the direction
of spin is opposite to the direction of quaternionic angular momentum.
\begin{spacing}{0.5}

\section{The momentum solutions of QRD equation}
\end{spacing}

To discuss the momentum like solutions of QRD equation (\ref{eq:42}),
we compare the quaternionic vector coefficient $e_{j}\,\left(\forall\,j=1,2,3\right)$
as,
\begin{align}
\left[\lambda D^{0}\left(A\right)L_{j}+D^{j}\left(A\right)E_{0}+\lambda\left(\overrightarrow{D}\times\overrightarrow{L}\right)_{j}-\mathscr{B}\lambda^{2}I_{j}\right]\boldsymbol{\Psi}= & \,\,0\:\,.\label{eq:66}
\end{align}
Now, putting the values of $D^{0}\left(A\right),$ $D^{j}\left(A\right)$
and $\mathscr{B}$ from equations (\ref{eq:37}) and (\ref{eq:38})
in given equation (\ref{eq:66}), we get
\begin{align}
\left(\begin{array}{cc}
\lambda\left(L_{j}-\lambda I_{j}\right) & i\left[e_{j}E_{0}+\lambda\left(\overrightarrow{e}\times\overrightarrow{L}\right)_{j}\right]\\
i\left[e_{j}E_{0}+\lambda\left(\overrightarrow{e}\times\overrightarrow{L}\right)_{j}\right] & \lambda\left(L_{j}+\lambda I_{j}\right)
\end{array}\right)\left(\begin{array}{c}
\Psi_{a}\\
\Psi_{b}
\end{array}\right)= & \,\,0\,\,,\label{eq:67}
\end{align}
which gives
\begin{align}
\lambda\left(L_{j}-\lambda I_{j}\right)\Psi_{a}+i\left[e_{j}E_{0}+\lambda\left(\overrightarrow{e}\times\overrightarrow{L}\right)_{j}\right]\Psi_{b}= & \,\,0\,\,,\label{eq:68}\\
\lambda\left(L_{j}+\lambda I_{j}\right)\Psi_{b}+i\left[e_{j}E_{0}+\lambda\left(\overrightarrow{e}\times\overrightarrow{L}\right)_{j}\right]\Psi_{a}= & \,\,0\,\,.\label{eq:69}
\end{align}
Equations (\ref{eq:68}) and (\ref{eq:69}) are identical to vector
analogy of Dirac's energy equations called generalized quaternionic
angular momentum equations which can describe respectively the angular
momentum of a electron and positron. Substituting the values of $\Psi_{a}$
and $\Psi_{b}$, we obtain the following equations:
\begin{align}
\Psi_{0}(E_{0},\overrightarrow{L})= & \,\,\frac{1}{i\lambda\left(L_{j}-\lambda I_{j}\right)}\left[e_{j}E_{0}+\lambda\left(\overrightarrow{e}\times\overrightarrow{L}\right)_{j}\right]\Psi_{2}(E_{0},\overrightarrow{L})\,\,,\label{eq:70}\\
\Psi_{1}(E_{0},\overrightarrow{L})= & \,\,\frac{1}{i\lambda\left(L_{j}-\lambda I_{j}\right)}\left[e_{j}E_{0}+\lambda\left(\overrightarrow{e}\times\overrightarrow{L}\right)_{j}\right]\Psi_{3}(E_{0},\overrightarrow{L})\,\,,\label{eq:71}\\
\Psi_{2}(E_{0},\overrightarrow{L})= & \,\,\frac{1}{i\lambda\left(L_{j}+\lambda I_{j}\right)}\left[e_{j}E_{0}+\lambda\left(\overrightarrow{e}\times\overrightarrow{L}\right)_{j}\right]\Psi_{0}(E_{0},\overrightarrow{L})\,\,,\label{eq:72}\\
\Psi_{3}(E_{0},\overrightarrow{L})= & \,\,\frac{1}{i\lambda\left(L_{j}+\lambda I_{j}\right)}\left[e_{j}E_{0}+\lambda\left(\overrightarrow{e}\times\overrightarrow{L}\right)_{j}\right]\Psi_{1}(E_{0},\overrightarrow{L})\,\,,\label{eq:73}
\end{align}
In the simplified form the above equations reduce to,
\begin{align}
\Psi_{\varLambda}(E_{0},\overrightarrow{L})+i\Omega_{+}(\overrightarrow{L})\left[e_{j}E_{0}+\lambda\left(\overrightarrow{e}\times\overrightarrow{L}\right)_{j}\right]\Psi_{\varLambda+2}(E_{0},\overrightarrow{L})= & 0\,,\,\left(\varLambda=0,1\right)\,\nonumber \\
\Psi_{\varLambda}(E_{0},\overrightarrow{L})+i\Omega_{-}(\overrightarrow{L})\left[e_{j}E_{0}+\lambda\left(\overrightarrow{e}\times\overrightarrow{L}\right)_{j}\right]\Psi_{\varLambda-2}(E_{0},\overrightarrow{L})= & 0\,,\,\left(\varLambda=2,3\right).\label{eq:74}
\end{align}
Here, $\Omega_{\pm}(\overrightarrow{L})=\frac{1}{\lambda\left(L_{j}\mp\lambda I_{j}\right)}$
is a rotational variable, can be used as $\Omega_{+}(\overrightarrow{L})$
for particle and $\Omega_{-}(\overrightarrow{L})$ for anti-particle
angular momentum. The term $\left(\overrightarrow{e}\times\overrightarrow{L}\right)$
shows a directional interaction between quaternion spin and orbital
angular momentum. Now, using equations (\ref{eq:70})-(\ref{eq:73})
we can analysis one, two and four component solutions for QRD equation.

\subsection{One component solutions}

We can start the angular momentum solutions with spin up state as
$\Psi_{0}=1$ and $\Psi_{1}=0$ and spin down states as $\Psi_{0}=0$
and $\Psi_{1}=1$ of Dirac-particle, and obtain
\begin{align}
\boldsymbol{\Psi}\rightarrow\boldsymbol{\Psi}^{\uparrow+}(E_{0},\overrightarrow{L})= & \,\,N_{+}^{L}\left(1-e_{2}\frac{i\left[e_{j}E_{0}+\lambda\left(\overrightarrow{e}\times\overrightarrow{L}\right)_{j}\right]}{\lambda\left(L_{j}+\lambda I_{j}\right)}\right)\,,\label{eq:75}\\
\boldsymbol{\Psi}\rightarrow\boldsymbol{\Psi}^{\downarrow+}(E_{0},\overrightarrow{L})= & \,\,N_{+}^{L}\left(e_{1}+e_{2}e_{1}\frac{i\left[e_{j}E_{0}+\lambda\left(\overrightarrow{e}\times\overrightarrow{L}\right)_{j}\right]}{\lambda\left(L_{j}+\lambda I_{j}\right)}\right)\,,\label{eq:76}
\end{align}
where $N_{+}^{L}=\,\,\frac{\lambda\left(L_{j}+\lambda I_{j}\right)}{\sqrt{\left[\lambda\left(L_{j}+\lambda I_{j}\right)\right]^{2}-\left[e_{j}E_{0}+\lambda\left(\overrightarrow{e}\times\overrightarrow{L}\right)_{j}\right]^{2}}}\,.$
Correspondingly, the angular momentum solutions for rotating Dirac
anti-particle, we take $\Psi_{2}=1$ and $\Psi_{3}=0$ for spin up
state and $\Psi_{2}=0$ and $\Psi_{3}=1$ for spin down states, then
\begin{align}
\boldsymbol{\Psi}\rightarrow\boldsymbol{\Psi}^{\uparrow-}(E_{0},\overrightarrow{L})= & \,\,N_{-}^{L}\left(\frac{-i\left[e_{j}E_{0}+\lambda\left(\overrightarrow{e}\times\overrightarrow{L}\right)_{j}\right]}{\lambda\left(L_{j}-\lambda I_{j}\right)}+e_{2}\right)\,,\label{eq:77}\\
\boldsymbol{\Psi}\rightarrow\boldsymbol{\Psi}^{\downarrow-}(E_{0},\overrightarrow{L})= & \,\,N_{-}^{L}\left(-e_{1}\frac{i\left[e_{j}E_{0}+\lambda\left(\overrightarrow{e}\times\overrightarrow{L}\right)_{j}\right]}{\lambda\left(L_{j}-\lambda I_{j}\right)}-e_{2}e_{1}\right)\:,\label{eq:78}
\end{align}
where $N_{-}^{L}=\,\,\frac{\lambda\left(L_{j}-\lambda I_{j}\right)}{\sqrt{\left[\lambda\left(L_{j}-\lambda I_{j}\right)\right]^{2}-\left[e_{j}E_{0}+\lambda\left(\overrightarrow{e}\times\overrightarrow{L}\right)_{j}\right]^{2}}}\,.$
Like energy solutions, the quaternionic angular momentum solutions
of one component describe the rotational momentum field for Dirac
particle and anti-particle.

\subsection{Two component solutions}

For the study of quaternionic two component angular momentum solutions,
we extant one component angular momentum solutions into two component
solution by using equation (\ref{eq:44}). Thus, the quaternionic
two component solutions corresponding to spin up and spin down states
of particle and anti-particle are expressed by,
\begin{align}
\boldsymbol{\Psi}\rightarrow\boldsymbol{\Psi}^{\uparrow+}(E_{0},\overrightarrow{L})= & \,\,\,N_{+}^{L}\left(\begin{array}{c}
1\\
\frac{-i\left[e_{j}E_{0}+\lambda\left(\overrightarrow{e}\times\overrightarrow{L}\right)_{j}\right]}{\lambda\left(L_{j}+\lambda I_{j}\right)}
\end{array}\right)\,,\label{eq:79}\\
\boldsymbol{\Psi}^{\uparrow-}(E_{0},\overrightarrow{L})= & \,\,\,\,N_{-}^{L}\left(\begin{array}{c}
\frac{-i\left[e_{j}E_{0}+\lambda\left(\overrightarrow{e}\times\overrightarrow{L}\right)_{j}\right]}{\lambda\left(L_{j}-\lambda I_{j}\right)}\\
1
\end{array}\right)\,,\label{eq:80}\\
\boldsymbol{\Psi}^{\downarrow+}(E_{0},\overrightarrow{L})= & -iN_{+}^{L}\left(\begin{array}{c}
\frac{i\left[e_{j}E_{0}+\lambda\left(\overrightarrow{e}\times\overrightarrow{L}\right)_{j}\right]}{\lambda\left(L_{j}+\lambda I_{j}\right)}\\
1
\end{array}\right)\,,\label{eq:81}\\
\boldsymbol{\Psi}^{\downarrow-}(E_{0},\overrightarrow{L})= & \,\,iN_{-}^{L}\left(\begin{array}{c}
1\\
\frac{i\left[e_{j}E_{0}+\lambda\left(\overrightarrow{e}\times\overrightarrow{L}\right)_{j}\right]}{\lambda\left(L_{j}-\lambda I_{j}\right)}
\end{array}\right)\,.\label{eq:82}
\end{align}
The two component solutions of quaternionic angular momentum show
that how spinors are associated with the rotational motion of particle
and anti-particle.

\subsection{Four component solutions}

Further, we also may write the four component solutions of quaternionic
angular momentum for spin up and spin down states of Dirac-particle
as,
\begin{align}
\boldsymbol{\Psi}\rightarrow\boldsymbol{\Psi}^{\uparrow+}(E_{0},\overrightarrow{L})\simeq & \,\,N_{+}^{L}\left(\begin{array}{c}
1\\
0\\
\frac{-i\left[e_{j}E_{0}+\lambda\left(\overrightarrow{e}\times\overrightarrow{L}\right)_{j}\right]}{\lambda\left(L_{j}+\lambda I_{j}\right)}\\
0
\end{array}\right)\,\,,\label{eq:83}\\
\boldsymbol{\Psi}^{\downarrow+}(E_{0},\overrightarrow{L})\simeq & \,\,N_{+}^{L}\left(\begin{array}{c}
0\\
1\\
0\\
\frac{i\left[e_{j}E_{0}+\lambda\left(\overrightarrow{e}\times\overrightarrow{L}\right)_{j}\right]}{\lambda\left(L_{j}+\lambda I_{j}\right)}
\end{array}\right)\,\,,\label{eq:84}
\end{align}
and for anti-particle, we obtain
\begin{align}
\boldsymbol{\Psi}\rightarrow\boldsymbol{\Psi}^{\uparrow-}(E_{0},\overrightarrow{L})\simeq & \,\,N_{-}^{L}\left(\begin{array}{c}
\frac{-i\left[e_{j}E_{0}+\lambda\left(\overrightarrow{e}\times\overrightarrow{L}\right)_{j}\right]}{\lambda\left(L_{j}-\lambda I_{j}\right)}\\
0\\
1\\
0
\end{array}\right)\,\,,\label{eq:85}\\
\boldsymbol{\Psi}^{\downarrow-}(E_{0},\overrightarrow{L})\simeq & \,\,\,N_{-}^{L}\left(\begin{array}{c}
0\\
\frac{i\left[e_{j}E_{0}+\lambda\left(\overrightarrow{e}\times\overrightarrow{L}\right)_{j}\right]}{\lambda\left(L_{j}-\lambda I_{j}\right)}\\
0\\
1
\end{array}\right)\,\,.\label{eq:86}
\end{align}
Thus, in quaternionic formalism, one, two and four component angular
momentum solutions are isomorphic to each other. The interesting part
in quaternionic description for Dirac equation is, it shows not only
rotational energy solution but also shows the rotational momentum
solutions for considering the four-dimensional Euclidean space-time.

\section{Quaternionic angular frequency and wave propagation vector}

We know that an electron rotate in a permissible orbit exhibits the
wave-nature. In order to calculate the angular frequency and wave
propagation vector of electron and positron in quaternionic space-time,
let us start with the quaternionic wave function $\boldsymbol{\Psi}$
consisting to $SU(2)$ group elements as \cite{key-42},
\begin{align}
\boldsymbol{\Psi}= & \,\,\stackrel[J=\frac{1}{2}]{\infty}{\Sigma}\left(2J+1\right)\stackrel[M=-J]{J}{\Sigma}T_{M}^{\mu J}\mathscr{D}_{S,M}^{J}\,\,\,\,\forall\,\left(\mu=0,1,2,3\right),\label{eq:87}
\end{align}
where $J,\,M$ and $S$ are denoted the total angular momentum, magnetic
quantum number due to total angular momentum and spin quantum number,
respectively. Here $\left(2J+1\right)$ defines the discrete value
of possible total angular momentum called the statistical weight,
$T_{M}^{\mu J}$ is a quaternionic variable associated with $SU(2)$
group and $\mathscr{D}_{S=\pm\frac{1}{2},M}^{J}$ is the Dirac spinor.
Thus, in quaternionic form, we have
\begin{align}
\Psi_{0}= & \,\,\stackrel[J=\frac{1}{2}]{\infty}{\Sigma}\left(2J+1\right)\stackrel[M=-J]{J}{\Sigma}T_{M}^{0J}\mathscr{D}_{\frac{1}{2},M}^{J}\,\,,\,\,\,\,\,\,\,\,\,(\text{corresponding to }e_{0})\label{eq:88}\\
\Psi_{1}= & \,\,\stackrel[J=\frac{1}{2}]{\infty}{\Sigma}\left(2J+1\right)\stackrel[M=-J]{J}{\Sigma}T_{M}^{1J}\mathscr{D}_{\frac{-1}{2},M}^{J}\,\,,\,\,\,\,\,(\text{corresponding to }e_{1})\label{eq:89}\\
\Psi_{2}= & \,\,\stackrel[J=\frac{1}{2}]{\infty}{\Sigma}\left(2J+1\right)\stackrel[M=-J]{J}{\Sigma}T_{M}^{2J}\mathscr{D}_{\frac{1}{2},M}^{J}\,\,,\,\,\,\,\,\,\,\,(\text{corresponding to }e_{2})\label{eq:90}\\
\Psi_{3}= & \,\,\stackrel[J=\frac{1}{2}]{\infty}{\Sigma}\left(2J+1\right)\stackrel[M=-J]{J}{\Sigma}T_{M}^{3J}\mathscr{D}_{\frac{-1}{2},M}^{J}\,\,,\,\,\,\,\,(\text{corresponding to }e_{3})\,.\label{eq:91}
\end{align}
Using above quaternionic rotational wave functions on equations (\ref{eq:50})
- (\ref{eq:53}), we obtain
\begin{align}
i\hslash\dot{T}_{M}^{0J}-\lambda^{2}I_{0}T_{M}^{0J}-i\lambda\left(\overrightarrow{e}\cdot\overrightarrow{L}\right)T_{M}^{2J}= & \,\,0\,,\label{eq:92}\\
i\hslash\dot{T}_{M}^{1J}-\lambda^{2}I_{0}T_{M}^{1J}-i\lambda\left(\overrightarrow{e}\cdot\overrightarrow{L}\right)T_{M}^{3J}= & \,\,0\,,\label{eq:93}\\
i\hslash\dot{T}_{M}^{2J}-\lambda^{2}I_{0}T_{M}^{2J}-i\lambda\left(\overrightarrow{e}\cdot\overrightarrow{L}\right)T_{M}^{0J}= & \,\,0\,,\label{eq:94}\\
i\hslash\dot{T}_{M}^{3J}-\lambda^{2}I_{0}T_{M}^{3J}-i\lambda\left(\overrightarrow{e}\cdot\overrightarrow{L}\right)T_{M}^{1J}= & \,\,0\,,\label{eq:95}
\end{align}
where we used energy operator $E_{0}\,\sim\,i\hslash\frac{\partial}{\partial t}$.
On the other hand, we may consider the general plane wave solution
of equation (\ref{eq:87}) in terms of quaternionic form as,
\begin{align}
T_{M}^{\mu J}= & \,\,G_{M}^{\mu J}\exp\left[-\frac{i}{\hslash}\left(\mathbb{P}\circ\mathbb{R}\right)\right]\,\,,\label{eq:96}
\end{align}
Here $G_{M}^{\mu J}$ is a constant, $\mathbb{P}$ and $\mathbb{R}$
are usual quaternionic four-momentum and four-space, respectively.
The scalar and vector components of quaternionic equation (\ref{eq:96})
are
\begin{align}
T_{M}^{0J}= & \,\,G_{M}^{0J}\exp\left[-\frac{i}{\hslash}\left(E_{0}t-\overrightarrow{p}\cdot\overrightarrow{r}\right)\right]\,,\,\,\,\hfill\,\,\,\text{\ensuremath{\left(\text{Coefficient of \ensuremath{e_{0}}}\right)}}\,\label{eq:97}\\
T_{M}^{aJ}= & \,\,G_{M}^{aJ}\exp\left[-\frac{i}{\hslash}\left(ct\overrightarrow{p}+\frac{E_{0}}{c}\overrightarrow{r}-\left(\overrightarrow{p}\times\overrightarrow{r}\right)\right)\right]\,\,,\,\,\,\,\text{\ensuremath{\left(\text{Coefficient of \ensuremath{e_{a}}}\right)}}\,\label{eq:98}
\end{align}
where $\ensuremath{a=1,2,3}.$ Now, substituting equations (\ref{eq:97})
and (\ref{eq:98}) and their time derivative in equations (\ref{eq:92})
- (\ref{eq:95}), we found
\begin{align}
\left(-\hslash\omega-\lambda^{2}I_{0}\right)G_{M}^{0J}-i\lambda\left(\overrightarrow{e}\cdot\overrightarrow{L}\right)G_{M}^{2J}= & \,\,0\,,\label{eq:99}\\
\left(-\hslash\omega-\lambda^{2}I_{0}\right)G_{M}^{1J}-i\lambda\left(\overrightarrow{e}\cdot\overrightarrow{L}\right)G_{M}^{3J}= & \,\,0\,\,,\label{eq:100}\\
\left(-\hslash\omega+\lambda^{2}I_{0}\right)G_{M}^{2J}-i\lambda\left(\overrightarrow{e}\cdot\overrightarrow{L}\right)G_{M}^{0J}= & \,\,0\,\,,\label{eq:101}\\
\left(-\hslash\omega+\lambda^{2}I_{0}\right)G^{3J}-i\lambda\left(\overrightarrow{e}\cdot\overrightarrow{L}\right)G_{M}^{1J}= & \,\,0\,\,.\label{eq:102}
\end{align}
Equations (\ref{eq:99}) and (\ref{eq:100}) are manifested to negative
energy solution while equations (\ref{eq:101}) and (\ref{eq:102})
are manifested to positive energy of Dirac particle. These dual-energy
equations can be reduced in $4\times4$ matrix form as,
\begin{align}
\left(\begin{array}{cccc}
-\hslash\omega-\lambda^{2}I_{0} & 0 & -i\lambda\left(\overrightarrow{e}\cdot\overrightarrow{L}\right) & 0\\
0 & -\hslash\omega-\lambda^{2}I_{0} & 0 & -i\lambda\left(\overrightarrow{e}\cdot\overrightarrow{L}\right)\\
-i\lambda\left(\overrightarrow{e}\cdot\overrightarrow{L}\right) & 0 & -\hslash\omega+\lambda^{2}I_{0} & 0\\
0 & -i\lambda\left(\overrightarrow{e}\cdot\overrightarrow{L}\right) & 0 & -\hslash\omega+\lambda^{2}I_{0}
\end{array}\right)\left(\begin{array}{c}
G_{M}^{0J}\\
G_{M}^{1J}\\
G_{M}^{2J}\\
G_{M}^{3J}
\end{array}\right)= & \,\,0\,,\label{eq:103}
\end{align}
which gives
\begin{align}
\omega\,:\longrightarrow\,\omega_{\pm}= & \pm\frac{\lambda\sqrt{\lambda^{2}I_{0}^{2}-\hslash^{2}\left(\overrightarrow{e}\cdot\overrightarrow{\nabla}_{\Theta}\right)^{2}}}{\hslash}\,\,.\label{eq:104}
\end{align}
Here, we used angular momentum $\overrightarrow{L}\rightarrow-i\hslash\overrightarrow{\nabla}_{\Theta},$
where $\overrightarrow{\nabla}_{\Theta}$ shows the rotational analog
of nabla operator corresponding to Euler angles $\left(\theta,\phi,\psi\right)$\cite{key-45}.
The term $\omega_{+}$ represented the angular frequency of the Dirac
like particle while the $\omega_{-}$ represented the angular frequency
of the its anti-particle. Similarly, for the general solution of angular
momentum equations, we substitute equations (\ref{eq:97}) and (\ref{eq:98})
in equations (\ref{eq:70}) - (\ref{eq:73}), and obtain
\begin{align}
\lambda\left(L_{j}-\lambda I_{j}\right)G_{M}^{0J}+i\left[e_{j}\left(c\overrightarrow{k}\right)+\lambda\left(\overrightarrow{e}\times\overrightarrow{L}\right)_{j}\right]G_{M}^{2J}= & 0\,\,,\label{eq:105}\\
\lambda\left(L_{j}-\lambda I_{j}\right)G_{M}^{1J}+i\left[e_{j}\left(c\overrightarrow{k}\right)+\lambda\left(\overrightarrow{e}\times\overrightarrow{L}\right)_{j}\right]G_{M}^{3J}= & 0\,\,,\label{eq:106}\\
\lambda\left(L_{j}+\lambda I_{j}\right)G_{M}^{2J}+i\left[e_{j}\left(c\overrightarrow{k}\right)+\lambda\left(\overrightarrow{e}\times\overrightarrow{L}\right)_{j}\right]G_{M}^{0J}= & 0\,\,,\label{eq:107}\\
\lambda\left(L_{j}+\lambda I_{j}\right)G^{3J}+i\left[e_{j}\left(c\overrightarrow{k}\right)+\lambda\left(\overrightarrow{e}\times\overrightarrow{L}\right)_{j}\right]G_{M}^{1J}= & 0\,\,.\label{eq:108}
\end{align}
where the wave propagation vector $\overrightarrow{k}\,\sim\,\frac{\overrightarrow{p}}{\hslash}$.
These equations also can be written in $4\times4$ matrix form as,
\begin{align}
\left(\begin{array}{cccc}
A_{-} & 0 & B & 0\\
0 & A_{-} & 0 & B\\
B & 0 & A_{+} & 0\\
0 & B & 0 & A_{+}
\end{array}\right) & \left(\begin{array}{c}
G_{M}^{0J}\\
G_{M}^{1J}\\
G_{M}^{2J}\\
G_{M}^{3J}
\end{array}\right)\,=\,\,0\,\,,\label{eq:109}
\end{align}
along with
\begin{align}
A_{\pm}\,= & \:\lambda\left(L_{j}\pm\lambda I_{j}\right)\,,\,\,B\,=\,i\left[e_{j}\left(c\overrightarrow{k}\right)+\lambda\left(\overrightarrow{e}\times\overrightarrow{L}\right)_{j}\right]\,.\label{eq:110}
\end{align}
Therefore, from equation (\ref{eq:109}) we obtain
\begin{align}
\overrightarrow{k}\,\,:\longrightarrow\,\,\overrightarrow{k}_{\pm}= & \,\,\pm\frac{ie_{j}\lambda\sqrt{\left(-\hslash^{2}\nabla_{\Theta_{j}}^{2}-\lambda^{2}I_{j}^{2}\right)}+i\hslash\lambda\left(\overrightarrow{e}\times\overrightarrow{\nabla}_{\Theta}\right)_{j}}{c}\,\,,\,\,\left(\forall\,j=1,2,3\right)\,.\label{eq:111}
\end{align}
Equation (\ref{eq:111}) represented an expression for wave vector
$\left(\overrightarrow{k}\right)$ corresponding to quaternionic angular
momentum that propagates along quaternionic basis $e_{j}$. Accordingly,
$\overrightarrow{k}_{+}$ can represent the wave propagation for Dirac
like particle while $\overrightarrow{k}_{-}$ can represent the wave
propagation corresponding to the anti-particle. Here, we should be
notice that $(\omega,\,\overrightarrow{k})$ shows the quaternionic
four-wave vector for Euclidean space.

\section{Conclusion}

In the present work, the generalized Dirac equation for rotating particle
has been demonstrated in term of quaternionic division algebra. Split
quaternion is the another variety of quaternion. The interesting part
of split quaternion is that, it can use not only the wave-mechanism
(Schr�dinger theory) but also use for matrix-mechanism (Heisenberg
theory) of quantum formalism. The matrix realization of split quaternion
shows the Pauli's spin state of fermions or anti-fermions. Therefore,
to visualizing the rotational properties of Dirac-like particles split
quaternions or quaternions algebra can be used for Euclidean space-time.\textcolor{blue}{{}
}In quaternionic field, we have discussed four angular momentum consisted
rotational energy and rotational momentum of an electron in four dimensional
Euclidean space-time. The connection between rotational matrices (tau-matrices)
along with quaternionic basis $\left(e_{0},\,e_{1},\,e_{2},\,e_{3}\right)$
has been described. Further, the quaternionic four-masses have been
associated with the rest mass corresponding to quaternionic scalar
basis and the moving mass corresponding to quaternionic vector basis
given by equation (\ref{eq:21}). We have also written the quaternionic
relativistic four-space, moment of inertia and angular momentum. The
components of quaternionic resultant rotational energy and rotational
momentum are established in compact and simple manner given by equations
(\ref{eq:28}) and (\ref{eq:29}). We have investigated a new form
of QRD equation (\ref{eq:42}) that unifies the rotational form of
Dirac-energy and Dirac- angular momentum of a particle in a single
framework. The solutions of QRD equation has been represented in one,
two and four components form which described the rotational motion
of Dirac particle and anti-particle with spin up and down states.
Further, we also have discussed a general form of quaternionic wave
function and its plane wave solutions in terms of quaternionic field.
The components of rotational energy solution of quaternionic wave
function leads to rotational frequency (\ref{eq:104}) while the components
of rotational momentum solution of quaternionic wave leads to wave
propagation vector (\ref{eq:111}) for Dirac spin-1/2 particles (or
anti-particles). Interestingly, the QRD equation for the rotating
particles also fulfills the conservation of quaternionic four rotational-momentum.
Therefore, we can conclude that the benefit of quaternionic formalism
is very important to study the rotating quantum particles as fermions
in four-dimensional Euclidean space-time. This theory can be extending
to discuss the behavior of rotating other subatomic particles in terms
of quaternionic field.

\end{document}